\documentclass[11pt,showpacs,superscriptaddress,floatfix]{revtex4-1}
\usepackage{amsmath,amssymb,amsfonts}
\usepackage{graphicx,epsfig}
\usepackage{caption}
\usepackage{xcolor}
\usepackage{times}

\begin{document}
\title{First clear evidence of quantum chaos in the bound states of an atomic nucleus}
\author{L. Mu\~noz}
\email{lmunoz@ucm.es}
\affiliation{Grupo de F\'{\i}sica Nuclear, Facultad de Ciencias F\'{\i}sicas, Universidad Complutense, E-28040 Madrid, Spain}
\author{R. A. Molina}
\affiliation{Instituto de Estructura de la Materia, IEM-CSIC, Serrano 123, Madrid, E-28006, Spain}
\author{J. M. G. G\'omez}
\affiliation{Grupo de F\'{\i}sica Nuclear, Facultad de Ciencias F\'{\i}sicas, Universidad Complutense, E-28040 Madrid, Spain}
\author{A. Heusler}
\affiliation{Gustav-Kirchhoff-Str. 7/1, D-69120 Heidelberg, Germany}

\begin{abstract}
We study the spectral fluctuations of the $^{208}$Pb nucleus using the complete experimental spectrum of 151 states up to excitation energies of $6.20$ MeV recently identified at the Maier-Leibnitz-Laboratorium at Garching, Germany. For natural parity states the results are very close to the predictions of Random Matrix Theory (RMT) for the nearest-neighbor spacing distribution. A quantitative estimate of the agreement is given by the Brody parameter $\omega$, which takes the value $\omega=0$ for regular systems and $\omega \simeq 1$ for chaotic systems. We obtain $\omega=0.85 \pm 0.02$ which is, to our knowledge, the closest value to chaos ever observed in experimental bound states of nuclei. By contrast, the results for unnatural parity states are far from RMT behavior. We interpret these results as a consequence of the strength of the residual interaction in $^{208}$Pb, which, according to experimental data, is much stronger for natural than for unnatural parity states. In addition our results show that chaotic and non-chaotic nuclear states coexist in the same energy region of the spectrum.
\end{abstract}


\maketitle

The atomic nucleus is generally considered a paradigmatic case of quantum chaos. Intuitively one can expect that fast moving nucleons interacting with the strong nuclear force and bound in the small nuclear volume should give rise to a chaotic motion. During the last three decades the quest for chaos in nuclei has been quite intensive, both with theoretical calculations using nuclear models and with detailed analyses of experimental data. Statistical spectroscopy studies in nuclei have been also motivated by a desire to understand the implications of chaotic behavior in many-body quantum systems. Theoretical calculations, especially shell-model calculations, have shown a strongly chaotic behavior of bound states at higher excitation energy, in regions of high level density. However, as we discuss below, it has not been possible up to now to observe chaos in the experimental bound energy levels of any single nucleus. For a comprehensive review of chaos in nuclei see for example G\'omez {\it et al.} \cite{Gomez11} and Weidenm\"uller and Mitchell \cite{Weidenmuller09}.

In this paper we analyze spectral fluctuations in $^{208}$Pb using the experimental data recently obtained by Heusler {\it et al.} \cite{Heusler16} from the study of $^{208}$Pb(p, p'), $^{207}$Pb(d, p) and \mbox{$^{208}$Pb(d, d')} reactions using the Q3D magnetic spectrograph at the Maier-Leibnitz-Laboratorium at Garching, Germany. There are 151 states at $E_x < 6.20$ MeV identified with spin and parity assignments and the $J^{\pi}$ level sequences are sufficiently long for reliable statistical analysis. We have found sequences of bound states with clearly chaotic statistics, while other sequences exhibit intermediate properties between chaos and regularity. These two different behaviors are related to the strength of the residual interaction that destroys the mean field order, as predicted by shell-model calculations. To our knowledge it is the first time that such a behavior is inferred directly from experimental nuclear states.

As is well known, the term quantum chaos refers strictly to quantum systems that are chaotic in the classical limit. For a system like the atomic nucleus, which has no classical limit, the term chaos started to be used when Haq, Pandey and Bohigas \cite{Haq82} analyzed the spectral fluctuations of a very large number of experimentally identified neutron and proton $J^{\pi} = 1/2^+$ resonances just above the one-nucleon emission threshold and showed that they agree very well with the spectral fluctuations of the Gaussian Orthogonal Ensemble (GOE) of Random Matrix Theory (RMT). According to the BGS conjecture \cite{Bohigas84} the agreement with GOE is characteristic of quantum chaos. On the contrary, spectral fluctuations that coincide with those of a Poisson distribution are characteristic of a regular quantum system \cite{Berry77}. 

Thus, in this sense, it is clear that nuclei are very chaotic in the energy region just above the one-nucleon emission threshold. But for bound states, the situation is not so clear, because a good analysis of fluctuations in experimental energy spectra requires the knowledge of sufficiently long, pure and complete sequences, i.e. with the same $J^{\pi}T$ values and without missing levels or $J^{\pi}T$ misassignments.
But this ideal situation is rarely found in nuclei. For very light nuclei the number of bound levels is not sufficient for statistical purposes. For medium and heavy nuclei the identified levels are limited to the ground state region, because at higher energy the level density becomes very high and the experimental identification of the energy and $J^{\pi}$ values becomes generally impossible.

Only in very few nuclei, namely $^{26}$Al and $^{30}$P, the full experimental spectrum has been essentially identified up to the proton separation energy at $E_x \simeq 8$ MeV. A statistical analysis of level fluctuations in these nuclei has been performed combining level spacings of different $J^{\pi}$ or $J^{\pi}T$ sequences \cite{Mitchell01}.  A surprising result is that the behavior of  the nearest neighbor spacing (NNS) distribution $P(s)$ and the behavior of the spectral rigidity statistic $\Delta_3(L)$ are quite similar in these nuclei whether or not the isospin $T$ is taken into account in the statistical analysis. As is well known, if a level sequence is not pure, i.e. contains states belonging to different symmetry classes, the strong level correlations characteristic of chaotic spectra are destroyed. Therefore one has to analyze separately the $J^{\pi}$ level sequences of different $T$. But both $^{26}$Al and $^{30}$P are odd-odd $N = Z$ nuclei, where states with isospin $T = 0$ and $T = 1$ have nearly equal density of levels in the ground state region, and an isospin symmetry breaking of about $3\%$ due to the Coulomb interaction may lead to similar fluctuation properties for $J^{\pi}$ or $J^{\pi}T$ sequences \cite{French88, Guhr90}. The isospin symmetry breaking seems to be responsible for the fact that the $P(s)$ statistic is equally far from GOE and Poisson predictions in these two nuclei.

Experimental pure level sequences in the ground-state region are generally too short for statistical analysis. But in order to improve statistics, level spacings from different nuclei can be combined into a single set to analyze the behavior of the NNS distribution $P(s)$. An extensive analysis of low-lying energy levels was performed by Shriner {\it et al.} \cite{Shriner91} using experimental data along the whole nuclear chart. A total of 988 spacings from 60 different nuclei were included in the analysis. A simple quantitative measure of chaos or regularity is provided by the Brody parameter $\omega$, which in the extreme cases takes the value $\omega \simeq 1$ for GOE and $\omega = 0$ for Poisson. For the whole set of 988 spacings the fit gives $\omega = 0.43 \pm 0.05$, which is an intermediate value closer to Poisson than to GOE. Separating the data in six different mass regions a clear trend from GOE to Poisson is observed as the nuclear mass increases. For light nuclei with $A \le 50$, the fit gives $\omega = 0.72 \pm 0.16$, and for the heaviest nuclei with $A > 230$ it gives $\omega = 0.24 \pm 0.11$. Generally spherical nuclei are closer to GOE and deformed nuclei are closer to Poisson. The latter is not necessarily a manifestation of regular behavior for the low-lying states of these nuclei, because a deviation towards Poisson may be also due to the omission of some symmetry. In the present case of deformed nuclei it may be due to omission of the $K$ quantum number, but it is not posible to quantify this effect with the available experimental data.

We may conclude that, for one reason or another, the analysis of fluctuations in experimental nuclear bound states has not shown the existence of clear chaotic motion, which should have a $P(s)$ distribution with $\omega$ close to 1. By contrast, several analyses of level fluctuations have shown almost regular nuclear dynamics in deformed nuclei, where collective motion is dominant. In spherical nuclei experimental energy levels exhibit an intermediate behavior between GOE and Poisson, although more chaotic than regular, especially in light nuclei.

As mentioned above, recent experiments \cite{Heusler16} using a Q3D magnetic spectrograph to study several particle exchange nuclear reactions have provided accurate data on the $^{208}$Pb level spectrum, with a typical resolution of 3 keV. Excitation energies are determined with a median uncertainty of 70 eV for $^{208}$Pb(p, p'). The study of (p, p') scattering via isobaric analog resonances in $^{209}$Bi provides the knowledge of up to 20 amplitudes of neutron 1p-1h configurations in each state. Essentially by this means the dominant structure, spin and parity for the 151 states at $E_x < 6.20$ MeV were identified.

Experimental nuclear spectra become increasingly plagued with unidentified states, missing levels and some misassignments as excitation energy increases. But we consider that these new accurate data on $^{208}$Pb enable a meaningful statistical analysis of level fluctuations in a nucleus, with pure, complete, and reasonably long sequences.

Fluctuations are the departure of the actual level density from a local uniform density. Therefore it is essential to eliminate the smooth part of the exponential increase of the nuclear density, mapping the actual spectrum onto a quasiuniform spectrum with mean spacing $\left< s \right> = 1$.
This step, called unfolding, is delicate and of utmost importance, because some of the unfolding procedures used in the literature can lead to completely wrong results on the behavior of level fluctuations \cite{Gomez02}. In this work we have used the constant temperature formula \cite{Shriner91},
\begin{equation}
\overline{\rho}(E) = \frac{1}{T} \exp[(E-E_0)/T],
\end{equation}
where $T$ and $E_0$ are taken as parameters, and separate unfolding has been performed for all $J^{\pi}$ sequences with a minimum of 5 known consecutive states. The longest sequence corresponds to the $J^{\pi} = 3^-$ states, with 19 consecutive spacings. Gathering the unfolded spacings for all $J^{\pi}$ into a single set, there are 115 spacings. 

\begin{figure*}
\vspace*{-0.5cm}
\begin{center}
\hspace*{-0.8cm}\rotatebox{0}{\scalebox{0.4}[0.4]{\includegraphics{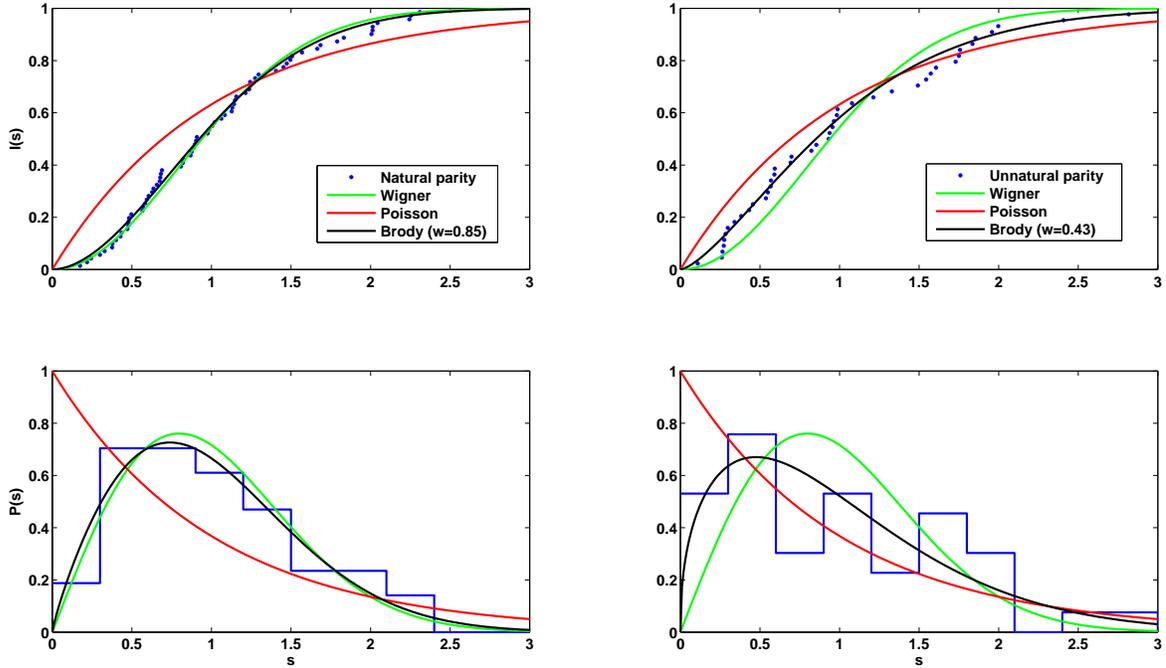}}}
\end{center}
\vspace*{-1cm}\caption{ (Color online). Nearest neighbor spacing (NNS) distribution $P(s)$ (down) and cumulative NNS distribution $I(s)$ (up) for the experimental states of $^{208}$Pb at $E_x < 6.20$ MeV, and the best fit Brody distribution, compared to the Wigner and Poisson distributions. Two set of states are considered: natural (left) and unnatural parity states (right).}
\label{nat_unnat_tot}
\end{figure*}

An assesment of chaotic or regular behavior is given by comparison of the NNS distribution to Poisson and Wigner.
The Poisson distribution is given by $P_P(s) = \exp(-s)$.
The Wigner surmise, $P_W(s) = (\pi s/2) \exp (-\pi s^2/4)$,
is a very good approximation to the GOE distribution.

An effective way to interpolate between the Poisson limit ($\omega = 0$) and the Wigner surmise ($\omega = 1$) is provided by the Brody distribution,
\begin{align}
& P_B(s,\omega) = (\omega+1) a_{\omega} s^{\omega} \exp(-a_{\omega} s^{\omega+1}),\\
& a_{\omega} = \left[\Gamma\left(\frac{\omega+2}{\omega+1}\right) \right]^{\omega+1}, \nonumber
\end{align}
where $\Gamma$ is the gamma function. The Brody parameter $\omega$ is given by the best fit to the histogram of $P(s)$. When the number of spacings is not very large, it is preferable to fit the cumulative distribution,
\begin{equation}
I(s) = \int_0^s P(x) dx.
\end{equation}

The cumulative Brody distribution is given by
\begin{equation}
I_B(s,\omega) = 1 - \exp(-a_{\omega} s^{\omega+1}).
\end{equation}

The fit to the full set of 115 experimental spacings gives $\omega = 0.63 \pm 0.08$ for $P(s)$ and $\omega = 0.68 \pm 0.02$ for $I(s)$. The two values are similar, but the cumulative fit is more accurate. Thus, from now on, we give the $\omega$ values for the cumulative distribution.

The relevant question with these results is how to interpret them regarding chaotic motion in $^{208}$Pb. Since it is a double closed-shell nucleus, the excited states in the ground-state region have a rather simple structure, dominated by one-particle one-hole (1p-1h) and a few 2p-2h configurations. In fact Heusler {\it et al.} \cite{Heusler16} have shown that the number of identified states at $E_x < 6.20$ MeV nearly agrees with the number of states in this energy interval predicted by what they call the ``extended schematic shell model'', which is a simplified shell-model consisting of 1p-1h mean field configurations plus the diagonal part of the surface delta interaction (SDI), extended with 2p-2h configurations with SDI. Clearly the basis configurations are spread out among a small number of states; often only two or three configurations are dominant in each state.
By contrast, GOE wave functions are not dominated by any particular amplitudes. Thus the similarities with GOE are certainly limited. The result $\omega = 0.68 \pm 0.02$ is similar to the value $\omega = 0.72 \pm 0.16$ for nuclei with $A<50$ \cite{Shriner91}. Therefore we may wonder whether it represents more or less a practical limit of possible chaos in nuclear bound states.

In the shell model the mean field gives rise to a regular motion and the residual interaction produces the mixing of basis states in the eigenstates, destroying the regular mean-field motion. This effect has been observed in shell-model calculations introducing a strength parameter to modulate the residual interaction. As the strength parameter increases, the fluctuations measures of energy levels approach GOE behavior and thus the motion becomes chaotic \cite{Zelevinsky96}.

Heusler {\it et al.} have shown \cite{Heusler16} that there is better agreement of the extended schematic shell model with experiment for unnatural parity states ($J^\pi= 0^-, 1^+, 2^-, \dots, 11^+ $) than for natural parity states ($J^\pi= 0^+, 1^-, 2^+, \dots, 12^+ $) in $^{208}$Pb. The excitation energies of 70 states with unnatural parity at $E_x < 6.20$ MeV agree within about 0.2 MeV with 1p-1h configurations of the extended schematic shell model. By contrast, the excitation energies of about 20 natural parity states are more than 0.5 MeV lower than the model prediction. Hence the residual interaction is much larger for natural than for unnatural parity states.

\begin{table}
\caption{Number of spacings and Brody parameter $\omega$ for the different combinations of parity in the experimental states of $^{208}$Pb at $E_x < 6.20$ MeV.}
\begin{tabular}{lccc}
\hline\hline
& & Number of spacings & \\ \cline{2-4}
Parity & all & natural & unnatural  \\
\hline
even & 45 & 29 & 16 \\
odd & 70 & 42 & 28 \\
all & 115 & 71 & 44 \\
\hline
Brody $\omega \;\;$ & $0.68 \pm 0.02$ & $0.85 \pm 0.02$ & $0.43 \pm 0.03$ \\
\hline
\hline
\end{tabular}
\label{spacings}
\end{table}

To check if this effect can be observed in the fluctuation measures, we have analyzed separately the NNS distribution of experimental natural and unnatural parity states. Table \ref{spacings} shows the number of unfolded spacings of each type at $E_x < 6.20$ MeV in $^{208}$Pb and the corresponding values of the Brody parameter obtained from the fit of the cumulative distribution. Fig. \ref{nat_unnat_tot} shows the distributions $I(s)$ and $P(s)$ for natural and unnatural parity states, compared to Wigner and Poisson. To guide the eye we have also plotted the Brody distribution.

For natural parity states the behavior is definitely chaotic, with $\omega = 0.85 \pm 0.02$. The Brody and Wigner curves for $P(s)$ nearly coincide, and the dots representing the cumulative distribution clearly follow the Wigner distribution. To our knowledge, this is the closest GOE behavior ever observed in experimental nuclear bound states. It is worth to comment that the Wigner surmise is only a good analytical approximation to the GOE distribution, and that the best fit of $P_B(s,\omega)$ to the exact $P(s)$ distribution for GOE is obtained for $\omega = 0.957$, not for $\omega = 1$ \cite{Brody81}.

By contrast, for unnatural parity states the cumulative distribution is intermediate between the two extremes, somewhat closer to Poisson, with $\omega = 0.43 \pm 0.03$. There are only 44 spacings and the histogram of $P(s)$ oscillates a lot. But notice especially the different behavior of $P(s)$ for small spacings in natural and unnatural parity states.
Level repulsion is seen to be much stronger for natural parity states. Strong level repulsion is characteristic of chaotic (Wigner-like) spectra, whereas for the regular motion $P(s)$ is maximal for small spacings.

We have also analyzed the NNS distributions for all the even and odd parity states. The Brody parameter is $\omega = 0.61 \pm 0.05$ for positive parity and $\omega = 0.67 \pm 0.04$ for negative parity. Hence, we do not observe any significant difference between even and odd parity.

As mentioned above, systematic studies of spectral fluctuations in experimental bound states throughout the nuclear chart have never found a clearly strong chaotic behavior. Shriner {\it et al.} \cite{Shriner91} obtain $\omega = 0.72 \pm 0.16$ for a set of 121 spacings grouping levels in nuclei with $A \le 50$. This is similar to our result $\omega = 0.68 \pm 0.02$ for the full set of states at $E_x < 6.20$ MeV in $^{208}$Pb. Let us point out that they obtain $\omega = 0.88 \pm 0.41$ for a set of 38 spacings in the mass region $50 < A \le 100$, but this result is quite uncertain. For $100 < A \le 150$ they obtain $\omega = 0.55 \pm 0.11$, and for heavier mass regions $\omega$ is always smaller, especially for deformed nuclei.

We should keep in mind that the experimental $J^{\pi}$ sequences in nuclei are generally quite short and that any missing levels or misassignments always bias the statistical measures towards Poisson. The same kind of bias is quickly produced by some broken symmetry or approximate symmetry in the nuclear states. Therefore intermediate results in the statistical fluctuation measures should be taken with caution. But in the case of $^{208}$Pb all the states with $E_x < 6.20$ MeV are now well identified, except for one or two tentative $J^{\pi}$ assignements at the upper end of the $5^+$, $7^+$ and $8^+$ states. We have checked that they essentially do not affect our results. Hence we are confident that the NNS distribution is not biased by missing levels or misassignments. The results reflect the degree of chaos caused by the residual interaction, and this is clearly seen comparing the NNS distributions of natural and unnatural parity states.

Let us briefly discuss now some theoretical results on chaos in nuclear bound states. Theoretical calculations provide long sequences of $J^{\pi}$ or $J^{\pi}T$ levels suitable for statistical analysis of fluctuations (no missing levels, no uncertain spin and parity assignments). Calculations performed with the spherical shell model, the cranking model, the interacting boson model and other models have shown examples of highly regular energy spectra in deformed nuclei and examples of highly chaotic spectra in spherical nuclei, although there are exceptions in some nuclei.

In spherical nuclei, where the shell model is most appropriate, many calculations have shown that for large configuration spaces, with $J^{\pi}T$ level sequences up to several thousand, the usual fluctuation measures $P(s)$ and $\Delta_3(L)$ agree very well with GOE predictions. To mention just some example, we highlight the work of Zelevinsky {\it et al.} \cite{Zelevinsky96} in $2s1d$ shell nuclei. In the middle of the $sd$ shell, $^{28}$Si has 12 valence nucleons and the $J^{\pi}T$ shell-model Hamiltonian matrices have large dimensionalities and the agreement with GOE is excellent.

In the $2p1f$ shell the configuration space and the level density are much larger than in $sd$-shell nuclei. Shell-model calculations with a realistic interaction have been performed to investigate the degree of chaos in different isotopes as a function of excitation energy by Molina {\it et al.} \cite{Molina00}. For example, in $^{46}$Sc a total of 25,498 spacings are included in the calculations, ensuring excellent statistics. The fluctuation statistics are in very good agreement with GOE, even for the low-lying levels above the yrast line, for all the Sc isotopes studied.
In $^{46}$Ti, an even-even nucleus, with lower density of states in the ground-state region, the agreement with GOE is excellent as well, even at low excitation energies.
However for Ca isotopes the results are quite different. The Brody parameter is always smaller in Ca than in Sc in all isotopes and energy regions, and also when the full spectrum is considered. Furthermore, at low energies the fluctuations are more regular than chaotic, for instance $\omega = 0.25$ for the levels up to 5 MeV above yrast in $^{52}$Ca. Similar results were obtained for Pb isotopes, with only valence neutrons outside the $^{208}$Pb core \cite{Molina06}.

These examples of shell-model calculations with a realistic interaction illustrate the same phenomenon that makes the difference in the chaotic vs. regular behavior of natural and unnatural parity states observed in the experimental energy levels of $^{208}$Pb. The shell-model residual nn interaction is much weaker than the residual pn interaction. Ca isotopes have only neutrons in the $pf$ shell, but if just one neutron is replaced by a proton, the pn interactions destroy the mean-field order. Therefore Sc or Ti isotopes, having both protons and neutrons in the valence space of the $pf$ shell, exhibit strong chaotic characteristics even in the ground-state region.

In conclusion, the recent identification with spin and parity assignment of all the 151 states at $E_x < 6.20$ MeV in $^{208}$Pb  by Heusler {\it et al.} \cite{Heusler16} has provided exceptionally long $J^{\pi}$ sequences of consecutive states free of missing levels and misassignments, enabling us to perform a reliable analysis of spectral fluctuations in this nucleus. Comparison of the experimental spectrum with extended 1p-1h schematic shell-model calculations clearly indicate that the residual interaction is much stronger for natural than for unnatural parity states \cite{Heusler16}. Therefore we have analyzed separately the spectral fluctuations of those two sets of states and have found that they behave very differently. The natural parity states exhibit results close to GOE and the unnatural parity states are far from GOE behavior. Thus these results clearly indicate that chaotic and non-chaotic states coexist in the energy region from the ground state up to 6.20 MeV excitation energy (the neutron threshold in $^{208}$Pb is $S$(n)$ = 7.368$ MeV). Furthermore, our analysis of the experimental spectrum has confirmed, to our knowledge for the first time, a well-known shell-model prediction, namely that chaos in nuclei arises when the residual interaction is strong enough to destroy the ordered motion of nucleons in the nuclear mean field.

\begin{acknowledgements}
This research has been conducted with support of the Spanish Ministry of
Science and Innovation grants FIS2012-35316 and FIS2012-34479.
\end{acknowledgements}

\end{document}